\documentclass[9pt, conference]{IEEEtran}
\IEEEoverridecommandlockouts
\usepackage{cite}

\usepackage{algorithmic}
\makeatletter
\renewcommand{\footnoterule}{%
    \kern -3pt 
    \hrule width 100pt height 0.4pt 
    \kern 2pt 
}
\makeatother
\usepackage{textcomp}
\usepackage[T1]{fontenc}
\usepackage{xcolor}
\usepackage{multirow, makecell, amssymb,amsfonts}
\usepackage[flushmargin]{footmisc}
\usepackage{amsmath,graphicx, array, caption, afterpage, adjustbox, subfig, lipsum, hyperref, endnotes}
\hypersetup{
    colorlinks=true,      
    linkcolor=black,       
    citecolor=black,       
    urlcolor=blue         
}

\begin{document}
\title{Wave-U-Mamba: An End-To-End Framework For High-Quality And Efficient Speech Super Resolution\\
}
\author{
\IEEEauthorblockN{Yongjoon Lee\textsuperscript{*}}
\IEEEauthorblockA{
\textit{Department of Statistics} \\
\textit{Korea University} \\
Seoul, South Korea \\
infected4098@korea.ac.kr
}
\and
\IEEEauthorblockN{Chanwoo Kim\textsuperscript{*}}
\IEEEauthorblockA{
\textit{Department of Artificial Intelligence} \\
\textit{Korea University} \\
Seoul, South Korea \\
chanwcom@korea.ac.kr
}
\thanks{*Corresponding authors: \{infected4098, chanwcom\}@korea.ac.kr}
}

\maketitle
\begin{abstract}
Speech Super-Resolution (SSR) is a task of enhancing low-resolution speech signals by restoring missing high-frequency components. Conventional approaches typically reconstruct log-mel features, followed by a vocoder that generates high-resolution speech in the waveform domain. However, as mel features lack phase information, this can result in performance degradation during the reconstruction phase. Motivated by recent advances with Selective State Spaces Models (SSMs), we propose a method, referred to as Wave-U-Mamba that directly performs SSR in time domain. In our comparative study, including models such as WSRGlow, NU-Wave 2, and AudioSR, Wave-U-Mamba demonstrates superior performance, achieving the lowest Log-Spectral Distance (LSD) across various low-resolution sampling rates, ranging from 8 to 24 kHz. Additionally, subjective human evaluations, scored using \text{Mean Opinion Score (MOS)} reveal that our method produces SSR with natural and human-like quality. Furthermore, Wave-U-Mamba achieves these results while generating high-resolution speech over nine times faster than baseline models on a single A100 GPU, with parameter sizes less than 2\% of those in the baseline models.
\end{abstract}
\vspace{-0.4cm}
\section{Introduction}
\vspace{-0.1cm}
\noindent Speech Super-Resolution (SSR), also referred to as Bandwidth Extension (BWE) is a process that aims to reconstruct High-Resolution (HR) speech from Low-Resolution (LR) speech. This field has gained significant attention due to the widespread availability of speech data recorded at low sampling rates, which can be attributed to hardware limitations, bandwidth constraints, and various other factors.

SSR is a generative task that leverages the power of modern generation backbones such as diffusion probabilistic models \cite{Han_2022, Lee_2021, liu2023audiosrversatileaudiosuperresolution}, Generative Adversarial Networks (GAN) \cite{Liu_2022, kumar2020nuganhighresolutionneural}, Generative Flow \cite{zhang2021wsrglowglowbasedwaveformgenerative}. One way to approach SSR involves generating HR waveform directly from LR waveform \cite{kuleshov2017audiosuperresolutionusing, sui2024trambahybridtransformermamba, tdBWE}. However, given that speech waveforms are characterized by significantly longer sequence lengths compared to mel spectrogram representations, conventional works first map the LR mel spectrogram to HR mel spectrogram and then project the recovered mel spectrogram onto a waveform through a well-defined vocoder \cite{liu2023audiosrversatileaudiosuperresolution, Liu_2022}. 
This two-step approach offers the advantage of decomposing the SSR task into simpler modules, akin to the traditional two-stage modeling employed in tasks such as Text-to-Speech (TTS) \cite{ren2022fastspeech2fasthighquality, shen2018naturalttssynthesisconditioning, wang2017tacotronendtoendspeechsynthesis}. 

Despite the promising results achieved by two-step approaches, certain limitations remain. The mel spectrogram, being a feature-reduced representation, inherently discards phase information during the conversion process. This introduces an additional challenge of implicit phase reconstruction during the SSR process. Previous works incorporate a pretrained vocoder to solve the problem so that SSR frameworks can only aim at recovering the high-frequency magnitudes. However, this approach often does not render fully end-to-end training of the model \cite{Liu_2022, liu2023audiosrversatileaudiosuperresolution, shen2018naturalttssynthesisconditioning} and is inefficient, as phase information could be preserved if the waveform were directly utilized.

Considering these factors, it is hypothesized that a model capable of directly generating HR waveforms from LR waveforms in an efficient and effective manner would offer a more optimal solution. In this paper, motivated by the success of selective SSMs \cite{gu2024mambalineartimesequencemodeling} to extract long-term dependent information both effectively and efficiently \cite{zhu2024visionmambaefficientvisual, lieber2024jambahybridtransformermambalanguage}, we propose Wave-U-Mamba to solve SSR in waveform domain. Wave-U-Mamba is a GAN-based framework that takes LR speech as input, with sampling rates ranging from 4 to 24 kHz, and upsamples it to HR speech with a target sampling rate of 48 kHz. Owing to the inherent capabilities of the Mamba architecture to efficiently process long sequences, coupled with its U-shaped model architecture that allows direct projection from LR to HR waveforms, the proposed model can generate HR waveforms much more rapidly than existing baseline models. Audio samples are made available online.\footnote{Listenable demos are available at \url{infected4098.github.io/waveumambademo}.}
\section{Related Works}
\subsection{\textit{Selective State Spaces Models}}\label{ssm}
\noindent Selective State Spaces Model, (commonly referred to as Mamba) offers significant advantages in capturing long-term dependencies when processing sequential data \cite{gu2024mambalineartimesequencemodeling}. Utilizing hardware-aware parallel scan algorithms, Mamba serves as a memory-efficient and high-performance feature extractor, capable of maintaining global dependencies across feature-dense sequences \cite{ma2024umambaenhancinglongrangedependency}. Given the considerable length of raw waveforms, computational complexity becomes a critical consideration in feature extraction. Mamba's near-linear computational complexity with respect to sequence length positions it as an optimal feature extraction method for waveform data. Integrating Mamba with U-net \cite{ronneberger2015unetconvolutionalnetworksbiomedical}  architecture has been explored in a variety of fields, including medical image segmentation \cite{ma2024umambaenhancinglongrangedependency, ruan2024vmunetvisionmambaunet}, image restoration \cite{deng2024cumambaselectivestatespace}.

\subsection{\textit{Time Domain Speech Processing}}
\noindent Although time-frequency representations such as mel spectrogram are traditionally employed to address speech-related tasks, time domain speech processing has been studied in various tasks including source separation \cite{luo2018tasnettimedomainaudioseparation, stoller2018waveunetmultiscaleneuralnetwork}, representation learning \cite{baevski2020wav2vec20frameworkselfsupervised}, multimodal processing \cite{akbari2021vatttransformersmultimodalselfsupervised}, speech generation \cite{oord2016wavenetgenerativemodelraw, goel2022itsrawaudiogeneration}, and speech super-resolution \cite{kuleshov2017audiosuperresolutionusing, sui2024trambahybridtransformermamba, tdBWE}. 

\begin{figure*}[!t]
\centering
\includegraphics[scale=0.29]{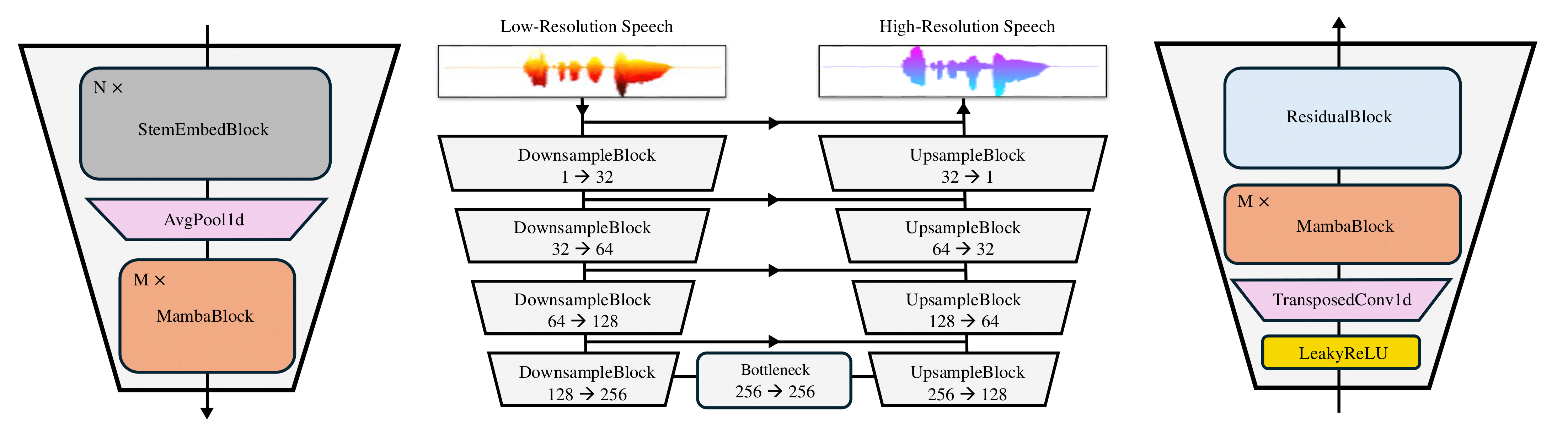}
\caption{The architecture of DownsampleBlock (Left), Wave-U-Mamba Generator (Middle), and UpsampleBlock (Right).}
\label{model_arc}
\end{figure*}

\vspace{-0.1cm}
\section{Wave-U-Mamba}

\noindent Wave-U-Mamba is a GAN-based generative model that comprises a single generator and two discriminators. Our generator is a U-Net based architecture that takes the raw waveform of LR speech as input and outputs HR speech waveform. A residual connection is employed every block to enhance stability. Especially, the residual block at the final upsampling layer enables the model to focus on estimating the upper-frequency components, given that the lower-frequency features are already captured in the LR waveform. Fig. \ref{model_arc} illustrates the overall architecture of the Wave-U-Mamba generator. Building on the previous work \cite{kong2020hifigangenerativeadversarialnetworks}, we utilize Multi-Period Discriminator (MPD) and Multi-Scale Discriminator (MSD), each comprising groups of sub-discriminators to perform adversarial training.

\subsection{\textit{Task Definition}}
\noindent
A LR speech signal has shape of $X_{L} = [x_{1, ... , s \times L}]$. $s$ represents signal length in seconds, and $L$ represents the sampling rate of the signal, where the upper-frequency limit is set to $L / 2$ by Nyquist Theorem. A HR signal has shape of $Y_{H} = [y_{1, ... , s\times H}]$. $H$ represents the sampling rate of the signal, where the upper-frequency limit is set to $H / 2$. We first upsample the LR signal to match the sampling rate of the HR signal (\textit{i.e.} $X_H = [x_{1, ... , s \times H}]$) using Fast Fourier Transform (FFT) interpolation. Then we parameterize and train the generative function $G$ so that $Y_H = G(X_H)$. 

\subsection{\textit{Generator}}
\begin{figure}[!t]
\centering
\includegraphics[scale=0.22]{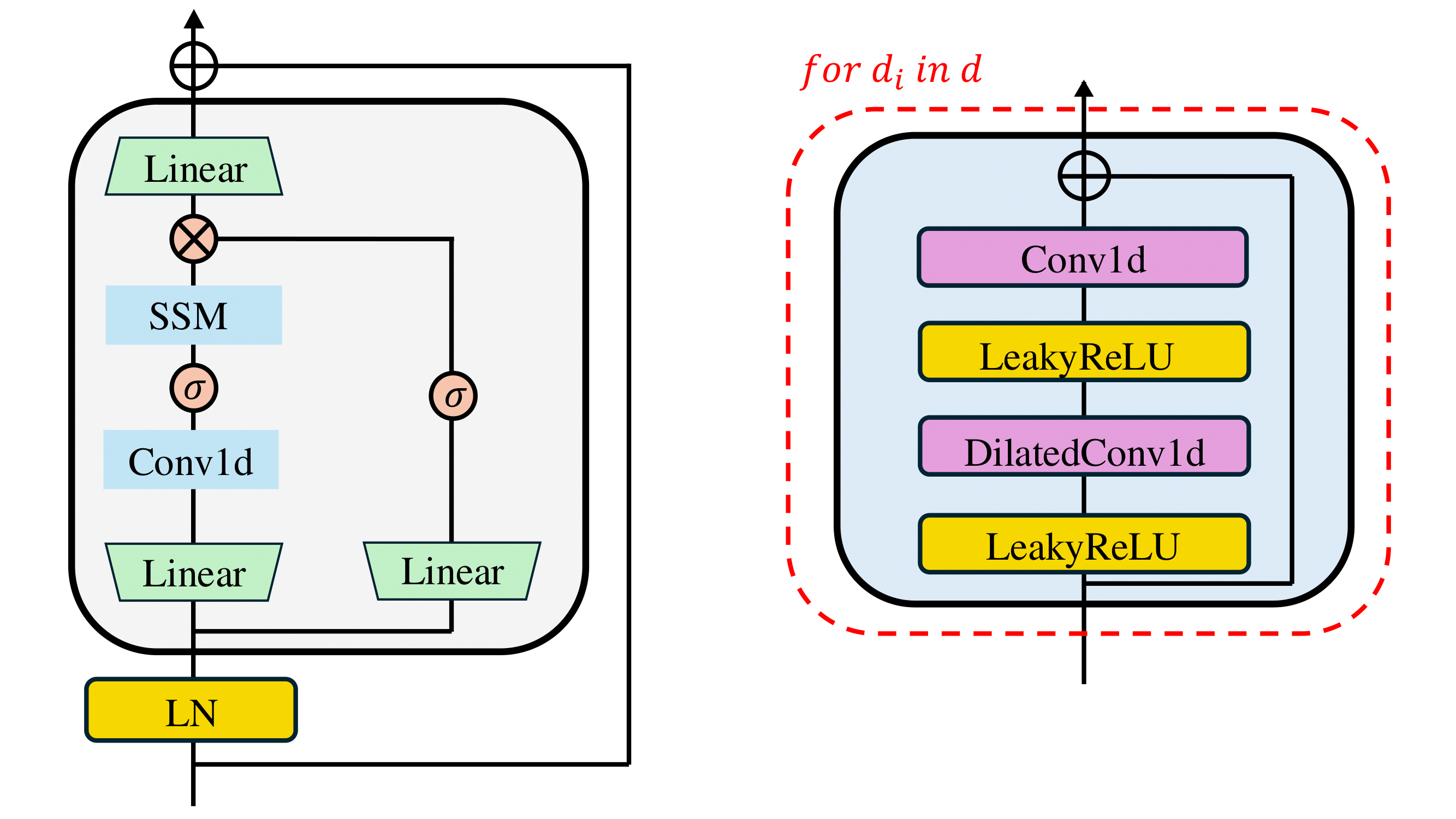}
\setlength{\abovecaptionskip}{10pt}

\caption{The architecture of MambaBlock (Left) and ResidualBlock (Right).}
\label{Details}
\end{figure}

\noindent\textbf{Downsampling}  Enhancing long-range dependency is a critical challenge in feature extraction within the waveform domain. Approaching distant features in waveform domain is harder than time-frequency domain due to the comparatively longer duration that a single frame in spectrogram spans. Previous studies have primarily focused on modifying the hyperparameters of convolutional neural networks (CNN) to extract long-term dependencies in waveform domain \cite{oord2016wavenetgenerativemodelraw, baevski2020wav2vec20frameworkselfsupervised}. However, these approaches often involve deepening network layers or expanding kernel sizes, leading to the loss of valuable information or increased computational complexity. To address this issue, we propose the incorporation of Mamba as a global feature extractor, due to its capability to efficiently and effectively learn long-range patterns. Specifically, we introduce MambaBlock that integrates Mamba with Layer Normalization (LN) \cite{ba2016layernormalization} and residual connection. We set $M$, the number of MambaBlocks to 2.

Mamba-based models are recognized for their difficulty in training, largely due to the relatively sharp or non-convex nature of their loss landscapes \cite{huang2024clipmambaclippretrainedmamba}. To mitigate this challenge, and motivated by \cite{xiao2021earlyconvolutionshelptransformers}, we design a 1D version of stem embedding as an operator to map input representations into a high-dimensional latent space, facilitating more stable training of the Mamba model. A StemEmbedBlock is made of 1D convolution with a kernel size of 4 and a stride of 1, followed by LN, LeakyReLU, and residual connection. In practice, we set $N$, the number of StemEmbedBlocks to 2 to double the feature dimension, which further stabilizes the training process. The hidden dimension at the bottleneck layer is set to 256, which is comparatively lower than that of other U-Net based SSR architectures \cite{Liu_2022, kuleshov2017audiosuperresolutionusing}. By constraining the hidden dimension size, the model effectively limits the propagation of extraneous patterns from the interpolated LR signal $X_H$, such as those generated by FFT interpolation, ensuring that only relevant and meaningful features are captured. 

As shown in previous work \cite{stoller2018waveunetmultiscaleneuralnetwork}, approaches such as strided CNNs with padding for sequence length reduction can often introduce borderline artifacts and disrupt the continuous nature of the signal due to the padding operation. To address the problem, we use average pooling to halve the sequence length. In this way, we compress features that add to the continuity of the generated HR signal. The bottleneck block follows the same structure as the DownsampleBlock but without pooling operation.

\noindent\textbf{Upsampling} We employ transposed convolution as an upsampling operator in which kernel sizes are set as a multiple of strides to reduce checkerboard artifacts common in transposed convolution \cite{odena2016deconvolution}. MambaBlocks with the same architecture in downsampling are utilized to capture global features. ResidualBlocks are put after Mambablocks to further improve the connectivity of representations and enhance the training stability. Specifically, every ResidualBlock has kernel sizes of 3 with dilation rates set to $d = [1, 3]$. The architecture of ResidualBlock can be found in Fig. \ref{Details}. Unlike other UpsampleBlocks, the final UpsampleBlock consists of a single convolutional layer, followed by a \textit{tanh} activation to predict the residual HR signal. All convolutional layers are weight-normalized \cite{salimans2016weightnormalizationsimplereparameterization}. 

\subsection{\textit{Discriminator}}
\noindent
Following early works on incorporating GAN for vocoding \cite{kong2020hifigangenerativeadversarialnetworks, kumar2019melgangenerativeadversarialnetworks}, we use Multi-Period Discriminator (MPD) and Multi-Scale Discriminator (MSD) as two discriminators to conduct adversarial training of Wave-U-Mamba. MPD first reshapes speech signal into 2D data and then applies 2D convolution to capture periodic patterns in the signal \cite{kong2020hifigangenerativeadversarialnetworks}. MPD consists of 5 sub-discriminators with different reshaping periods to reshape speech signals. MSD is a 1D convolution-based discriminator that operates on smoothed waveform \cite{kumar2019melgangenerativeadversarialnetworks}. MSD consists of 3 sub-discriminators that operate on raw waveform and two downsampled waveforms. By combining two modules, our generator is expected to capture both periodic patterns and continuous nature of speech signals, rendering natural and high-quality generation of HR speech.
Given that the discriminators in SSR tasks deal with comparatively fine-grained HR speech signal than those used in common vocoding tasks, we deepen the structure of both MPD and MSD by adding more building blocks.

\subsection{\textit{Training Objectives}}
\noindent\textbf{Mel Spectrogram Loss} For the natural and realistic SSR, it is crucial that perceptual quality be measured thoroughly. We therefore use {\it L1}-loss of the mel spectrogram of the predicted signal and the ground truth signal, where $\psi$ refers to the mel transformation. In equation \ref{mels}, $y$ and $x$ refer to the ground truth HR and the LR waveform, respectively. A function $G$ is our generator Wave-U-Mamba that takes as input $x$ to predict the HR waveform $y$.

\begin{equation}
L_{\text{mel}} = \mathbb{E}_{(y, x)}[||\psi(y) - \psi(G(x))||_1]
\label{mels}
\end{equation}
\noindent \textbf{Multi-Resolution Short-Time Fourier Transform (STFT) Loss} Since SSR is a task that aims to reconstruct high-frequency components, we incorporate Multi-Resolution Short-Time Fourier Transform (STFT) Loss \cite{yamamoto2020parallelwaveganfastwaveform} as another loss term for the model to learn detailed characteristics of high-frequency bands in the spectrogram. Multi-Resolution STFT Loss is the sum of the STFT losses with different parameters such as window size or hop length, where $M$ represents the number of STFT losses. Each STFT loss $L_{\mathrm{s}}$ is the sum of spectral convergence loss and log STFT magnitude loss. 

\vspace{-0.4cm}
\begin{equation}
\begin{aligned}
L_{\text{STFT}}(y, G(x)) &= \frac{1}{M}\sum^M_{m=1}L_{\text{s}}^{(m)}(y, G(x)) \\ 
L_s(y, G(x)) &= \mathbb{E}_{(y, x)}[L_{sc}(y, G(x)) + L_{mag}(y, G(x))]
\end{aligned}
\label{eq:multispec}
\end{equation}

\noindent \textbf{GAN Loss}
As originally illustrated in \cite{kong2020hifigangenerativeadversarialnetworks}, we use the same formulations for adversarial loss for both the generator and discriminator. 

\begin{equation}
\begin{aligned}
L_{\text{GAN}}(D;G) &= \mathbb{E}_{(y, x)}\left[(D(y)-1)^2 + \left(D(G(x))\right)^2\right] \\ 
L_{\text{GAN}}(G;D) &= \mathbb{E}_{x}\left[(D(G(x)) -1)^2\right]
\end{aligned}
\end{equation}

\noindent \textbf{Final Loss} 
Our final losses for the Wave-U-Mamba generator and discrminator are $L_{\text{G}}$ and $L_{\text{D}}$. $L_{\text{G}}$ is the weighted sum of Mel spectrogram loss, Multi-Resolution STFT loss, and the adversarial generator loss. 

\begin{equation}
\begin{aligned}
L_{\text{G}} &= \lambda_{\text{mel}}L_{\text{mel}} + \lambda_{\text{STFT}}L_{\text{STFT}} + L_{\text{GAN}}(G;D) \\
L_{\text{D}} &=  L_{\text{GAN}}(D;G)
\end{aligned}
\end{equation} 
We set $\lambda_{\text{mel}} = 45$ and $\lambda_{\text{STFT}}=10$. The coefficients were set to balance the overall scale of loss terms. We did not use {\it L1} Loss of the two waveforms because minimizing the distance between two waveforms does not direct to better perceptual quality nor the recovery of the high-frequency components, as pointed out in previous literature \cite{bwe, metricgan+}.

\section{Experimental Results}
\noindent
In this section, we describe the training settings and evaluation criteria used in our experiments. Then we visualize some sample mel spectrograms to help understand our results, which is followed by both objective and subjective score comparison of our models with other baselines, including WSRGlow \cite{zhang2021wsrglowglowbasedwaveformgenerative}, Nu-Wave 2 \cite{Han_2022} and AudioSR \cite{liu2023audiosrversatileaudiosuperresolution}. Furthermore, we show the results of ablation studies to demonstrate the characteristics of the design choices. Lastly, we show the efficiency measure of our model compared with other models. 

\subsection{\textit{Training}}
\noindent
VCTK \cite{vctk} is used as a dataset both for training and validation. The data were first cut into \textit{VCTK-train} and \textit{VCTK-test}, following the same data splitting scheme from the previous work \cite{kuleshov2017audiosuperresolutionusing} then were sampled at 48 kHz. We segmented about 0.7 seconds of data to use for training. We randomly set the cutoff frequency between 2 kHz and 12 kHz and then passed the data through an order-8 Chebyshev type 1 filter. The signal is downsampled to a target sampling rate and then upsampled back to 48 kHz to remove the high-frequency components completely. We normalized the signal onto $[-1, 1]$ in training because our model has \textit{tanh} activation on the last layer to generate the HR signal. For the calculation of the mel spectrogram, we use 80 mel filterbanks, the Hanning window with a window length of 2048 and a hop length of 512 for both the training and the evaluation. For the computation of Multi-Resolution STFT loss, we followed the default configuration of the open-source tool.\footnote{\url{https://github.com/csteinmetz1/auraloss}}

The model was trained for 25 epochs, during which we stopped the training when the validation loss does not decrease for three consecutive epochs. AdamW\cite{adamw} is used as an optimizer with $\beta_1=0.6$ and $\beta_2=0.99$. We set the learning rate to $4e-5$ for both the generator and the discriminator, which linearly increases to $2e-4$ in epoch 5 and then decreases with exponential weight decay with $\gamma$ set to 0.999. We apply gradient clipping to both the generator and the discriminator with a maximum norm of 2 to stabilize the training process. Experiments were conducted under 2 RTX A6000 GPUs with batch size set to 64, which took approximately 8 hours to train our model. 

\subsection{\textit{Evaluation}}
\noindent
The primary evaluation metric for the objective evaluation is Log-Spectral Distance (LSD), illustrated in Equation \ref{metrics-lsd}. $Y(f, t)$ and $\hat{Y}(f,t)$ refer to the magnitude spectrogram of the ground truth signal and the predicted signal. LSD can measure how well the signal has recovered the frequency components. 

\begin{equation}
    \label{metrics-lsd}
    \mathrm{LSD}(Y,\hat{Y}) = \frac{1}{T}\sum_{t=1}^{T}\sqrt{\frac{1}{F}\sum_{f=1}^{F}\log_{10}\left(\frac{Y(f,t)^2}{\hat{Y}(f,t)^2}\right)^2}
\end{equation}
\\
To evaluate the perceptual quality of the generated speech, we implemented 5-scale Mean Opinion Score (MOS) tests on samples from WSRGlow \cite{zhang2021wsrglowglowbasedwaveformgenerative}, Wave-U-Mamba, and AudioSR \cite{liu2023audiosrversatileaudiosuperresolution} then provided MOS scores. 45 raters were tested on 5 speech samples randomly selected from the \textit{VCTK-test} to measure the overall quality of the speech. The sampling rate of 8 kHz and 12 kHz were used for simulating the LR speech. All samples were anonymized and volume-normalized to guarantee the fairness of the testing procedure. 
 \begin{figure}[!t]
        \subfloat[]{%
            \includegraphics[width=.5\linewidth]{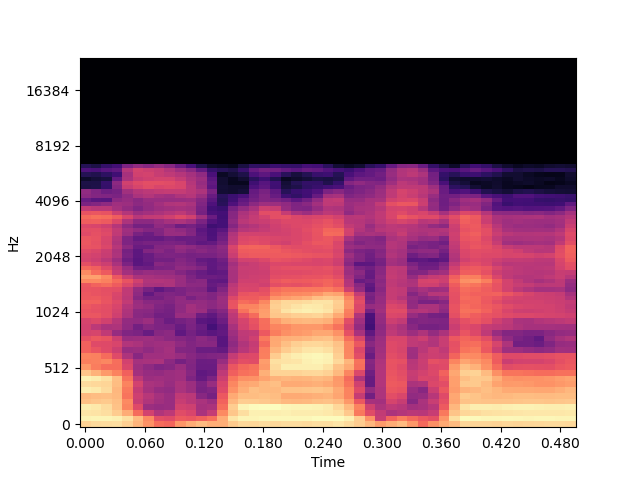}%
            \label{subfig:a}%
        }
        \subfloat[]{%
            \includegraphics[width=.5\linewidth]{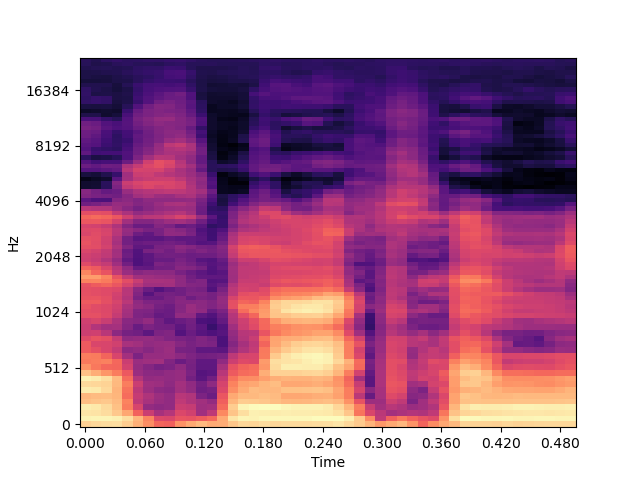}%
            \label{subfig:b}%
        } \\[0.1mm] 
        \subfloat[]{%
            \includegraphics[width=.5\linewidth]{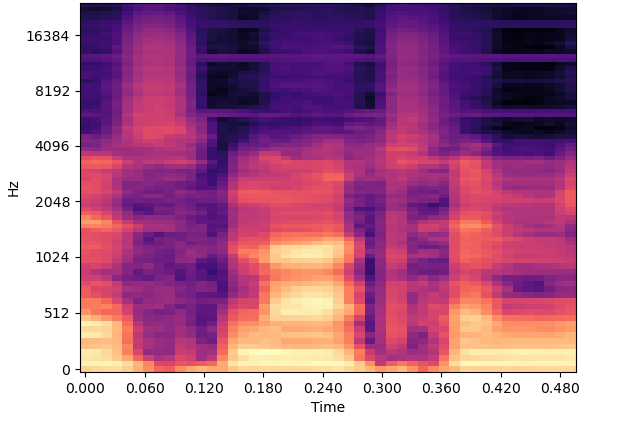}%
            \label{subfig:c}%
        }
        \subfloat[]{%
            \includegraphics[width=.5\linewidth]{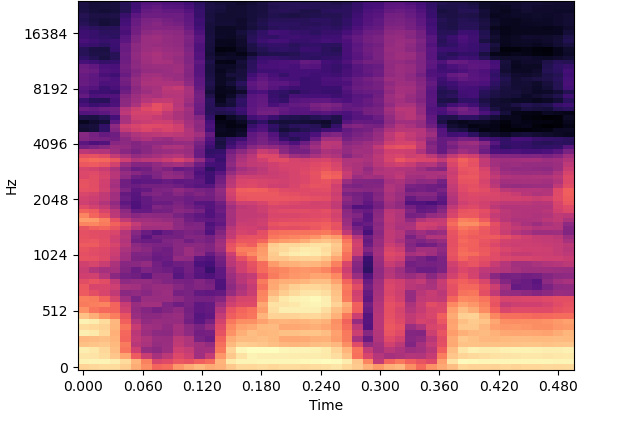}%
            \label{subfig:d}%
        }
        \caption{Enhanced mel spectrograms using Wave-U-Mamba at different training phases. \ref{subfig:a} and \ref{subfig:b} represent Low-Resolution and High-Resolution mel spectrogram of the ground truth speech signal. \ref{subfig:c} and \ref{subfig:d} represent recovered mel spectrogram from epoch 2 and 9, respectively. }
        \label{fig:fig}
    \label{exf}
    \end{figure}
    
\subsection{\textit{Results}}

\noindent Fig. \ref{exf} illustrates the sample mel spectrograms in training our model. We can see the checkerboard artifacts in Fig. \ref{subfig:c}, but by iterative training, artifacts are mitigated and high-frequency patterns are captured more precisely. This is consistent with the previous finding \cite{pons2021upsamplingartifactsneuralaudio} that long training could lead to the vanishing of checkerboard artifacts in many upsampling applications.

\begin{table}[!t]
\captionof{table}{{\textit{Comparison of Log-Spectral Distance (LSD) on various input sampling rates with target sampling rate of 48 kHz with other baseline models.}}}
\scalebox{1.25}{
 \begin{tabular}{*{6}{c}}
      \hline
      \makecell{Sampling Rate (kHz)} & 8 & 16 & 24 & AVG   \\
      \hline
  WSRGlow \cite{zhang2021wsrglowglowbasedwaveformgenerative}  & 1.05   & 0.84 & \textbf{0.70} & 0.86  \\
  \hline
  Nu-Wave 2 \cite{Han_2022} & 1.14 & 0.92 & 0.77 & 0.94 \\
  \hline
  AudioSR \cite{liu2023audiosrversatileaudiosuperresolution} & 1.30  & 1.11 & 0.94 & 1.12  \\
  \hline
  Wave-U-Mamba (Ours)  & \textbf{0.87} & \textbf{0.75} & \textbf{0.70} & \textbf{0.77} \\
  \hline
  \end{tabular}
}

  \label{LSDothers}
\end{table}

\noindent \textbf{Objective Performance Comparison} We have conducted an objective comparison of model performance with other State-Of-The-Art (SOTA) models using LSD.  Table \ref{LSDothers} shows that our model Wave-U-Mamba excels other models in terms of LSD, meaning that our model effectively captures high-frequency patterns. Additionally, our model exhibits a relatively consistent performance for audio chunks of any sampling rate. We can infer that this is due to the robustness of waveform representations, whereas for other models input mel representations can have varying amounts of active pixels.

\noindent \textbf{Subjective Performance Comparison} As seen in the table \ref{efficiency}, the result of the MOS suggest that our model has the best perceptual quality among all the other baseline models in improving LR speech signal to HR speech signal. In particular, our results have improved MOS scores of LR speech with a margin of 1.44. 

\noindent \textbf{Ablation studies} As presented in table \ref{ablations}, we conducted a series of experiments with various design configurations of the Wave-U-Mamba model and then compared the LSD to better understand the functioning of some of the key components of the model. Wave-U-Mamba-\textit{deep} refers to the same model architecture with the number of DownsampleBlock and UpsampleBlock increased to 5, which is set to 4 by default in Fig. \ref{model_arc}. The dimension size at the bottleneck layer is also doubled. The resulting LSD of  Wave-U-Mamba-\textit{deep} is consistent with the idea that projecting LR signal into a relatively low-dimensional space is sufficient for accurate reconstruction.
We also explored the impact of incorporating {\it L1} loss in the waveform domain in training. We observe that it not only deteriorated the performance but also introduced metallic artifacts in the reconstructed waveform. Lastly, removing MambaBlock from the model resulted in a marked degradation in performance, implying that Mamba is a crucial component of our model given that it functions as a global dependency extractor.

\begin{table}[!t]
\centering
\caption{\textit{Comparison of Mean Opinion Score (MOS), inference time, and the number of parameters with other baseline models. The evaluation of ground truth High-Resolution (HR) and Low-Resolution (LR) signals is included to demonstrate the margin of perceptual quality improvement achieved by Speech Super-Resolution (SSR) models.}}
\resizebox{\columnwidth}{!}{
\begin{tabular}{c|c|c|c}
\hline
Model & MOS $\uparrow$ & \makecell{Inference time \\ (ms)} $\downarrow$ & \makecell{\# Parameters \\ (M)} $\downarrow$ \\ 
\hline
WSRGlow \cite{zhang2021wsrglowglowbasedwaveformgenerative} & 3.28 & 45.2 $\pm$ 0.9 & 229.9 \\ 
\hline
AudioSR \cite{liu2023audiosrversatileaudiosuperresolution} & 3.27 & 2895.9 $\pm$ 11.7 & 500.0 \\ 
\hline
Wave-U-Mamba & \textbf{4.01} & \textbf{5.4 $\pm$ 0.9} & \textbf{4.2} \\ 
\Xhline{1.5pt} 
Ground Truth HR & 4.24 & - & - \\ 
\hline
Ground Truth LR & 2.57 & - & - \\ 
\hline
\end{tabular}
}
\label{efficiency}
\end{table}

\begin{table}[!t] 
\centering
\caption{\textit{Ablations of design choices of Wave-U-Mamba with regard to Log Spectral Distance (LSD) on various input sampling rates with target sampling rate of 48 kHz.}}
\resizebox{\columnwidth}{!}{ 
\begin{tabular}{*{7}{c}}
\hline
Sampling Rate (kHz) & 4  & 8  & 12  & 16  & 24 & AVG \\
\hline
Wave-U-Mamba & 0.97 & \textbf{0.87} & \textbf{0.81} & \textbf{0.75} & \textbf{0.70} & \textbf{0.82} \\
\hline
\makecell{Wave-U-Mamba \\ \textit{deep}} & \textbf{0.95} & 0.88 & 0.86 & 0.79 & 0.73 & 0.84 \\
\hline
\makecell{Wave-U-Mamba \\ \textit{with {\it L1} loss}} & 0.98 & 0.91 & 0.88 & 0.85 & 0.81 & 0.89 \\
\hline
\makecell{Wave-U-Mamba \\ \textit{without Mamba}} & 1.09 & 0.99 & 0.93 & 0.89 & 0.84 & 0.95 \\
\hline
\end{tabular}
}
\label{ablations} 
\end{table}

\noindent\textbf{Efficiency Comparison} To compare the efficiencies of the models, we used mean time to generate a second of speech fragment and the number of parameters as criteria. To calculate the inference speed, we first randomly sample files from VCTK, then estimate the mean inference time in milliseconds for generating one second of HR speech using a Monte Carlo simulation with 50 iterations. Inference speed is measured using a single A100 GPU. As indicated in Table \ref{efficiency}, our model contains fewer than 2\% of the parameters of the baseline models and exhibits an inference speech 536.3 times faster than AudioSR.

\section{Conclusion}
\noindent
In this work, we present Wave-U-Mamba, a fully end-to-end speech super-resolution model that operates in the waveform domain. The proposed model employs a U-structured architecture with generative adversarial networks for training. Wave-U-Mamba demonstrates the state-of-the-art performance on \textit{VCTK-test} dataset, surpassing existing methods without relying on additional models or pretrained weights. Furthermore, the model is capable of upsampling speech at any input resolution, from a sampling rate of 4 to 24 kHz, to a high-resolution output at 48 kHz, consistently delivering high-quality results. The model's efficiency in upsampling low-resolution speech signals is mainly attributed to its simple architecture, allowing for both computational efficiency and superior performance.

\section{Acknowledgements}
\vspace{-0.1cm}
\noindent
This work was partly supported by the Institute of Information \& Communications Technology Planning \& Evaluation(IITP)-ITRC(Information Technology Research Center) grant funded by the Korea government(MSIT)(IITP-2025-RS-2024-00436857); Institute of Information \& communications Technology Planning \& Evaluation (IITP) grantfunded by the Korea government(MNIST)(No. RS-2019-||190079.Artificial Intelligence Graduate School Program(Korea University))

\newpage
\begingroup
\renewcommand{\baselinestretch}{1.1} 
\large 
\bibliographystyle{IEEEbib}
\bibliography{strings,refs}
\endgroup

\end{document}